**Deep Learning-Based Prediction of PET Amyloid Status Using Multi-Contrast MRI**

Donghoon Kim[1], Jon André Ottesen[1,2,3], Ashwin Kumar[1], Brandon C. Ho[1], Elsa Bismuth[1], Christina B. Young[4], Elizabeth Mormino[4], and Greg Zaharchuk[1], for the Alzheimer's Disease Neuroimaging Initiative[5]

1. Department of Radiology, Stanford University, Stanford, USA
2. Computational Radiology & Artificial Intelligence (CRAI) Research Group, Division of Radiology and Nuclear Medicine, Oslo University Hospital, Oslo, Norway
3. Department of Physics, Faculty of Mathematics and Natural Sciences, University of Oslo, Oslo, Norway
4. Department of Neurology and Neurological Sciences, Stanford University, Stanford, USA
5. Alzheimer's Disease Neuroimaging Initiative investigators are listed at https://adni.loni.usc.edu/wp-content/uploads/how_to_apply/ADNI_Acknowledgement_List.pdf.

**Corresponding author**
Greg Zaharchuk
Department of Radiology, Stanford University, Stanford, CA, USA
gregz@stanford.edu

Donghoon Kim
Department of Radiology, Stanford University, Stanford, CA, USA
dknkim@stanford.edu

# Abstract

**Background**
Identifying amyloid-beta (Aβ) positive patients is critical for determining eligibility for Alzheimer's disease (AD) clinical trials and new disease-modifying treatments, but this is currently only possible with PET or CSF sampling. Previously reported MRI-based deep learning models for predicting amyloid positivity have used only T1-weighted (T1w) sequences and have shown moderate performance.

**Purpose**
To train deep learning models to predict amyloid PET positivity and to determine whether multi-contrast inputs can improve performance.

**Materials and Methods**
A total of 4,058 exams with multi-contrast MRI and PET-based quantitative Aβ deposition were obtained from three separate public datasets: the Alzheimer's Disease Neuroimaging Initiative (ADNI), the Open Access Series of Imaging Studies 3 (OASIS3), and the Anti-Amyloid Treatment in Asymptomatic Alzheimer's Disease (A4). Aβ positivity was defined based on each dataset's recommended centiloid threshold, and 55% of the cohort were amyloid positive. Two separate EfficientNet models were trained for amyloid positivity prediction: one with only T1w images and the other with both T1w and T2-FLAIR images as network inputs. The area under the curve (AUC), accuracy, sensitivity, and specificity were determined using an internal held-out test set. The trained models were further evaluated using an external test set. DeLong's and McNemar's tests were used to assess AUC and accuracy, respectively. Finally, we compared our results with a publicly available Aβ deep learning-based prediction model for T1w imaging.

**Results**
In the held-out test sets, the T1w and T1w+T2-FLAIR models demonstrated AUCs of 0.62 (95% CI: 0.60, 0.64) and 0.67 (95% CI: 0.64, 0.70) ($p = 0.006$); accuracies were 61% (95% CI: 60%, 63%) and 64% (95% CI: 62%, 66%) ($p = 0.008$); sensitivities were 0.88 and 0.71; and specificities were 0.23 and 0.53, respectively. The trained models showed similar performance in the external test set. Performance of the current model on both test sets exceeded that of the publicly available model.

**Conclusion**
The use of multi-contrast MRI, specifically incorporating T2-FLAIR in addition to T1w images, significantly improved the predictive accuracy of PET-determined amyloid status from MRI scans using a deep learning approach.

**Introduction**

Alzheimer's disease (AD) is a progressive disorder with three phases: an asymptomatic phase, mild cognitive impairment (MCI), and a dementia phase. The early detection of neuropathological changes in AD is valuable as it potentially enables earlier treatment (1). AD is commonly defined by its neuropathological hallmarks. Among these, amyloid-beta (Aβ) deposition is one of the earliest indicators of disease. In many clinical trials for AD, Amyloid deposition has been treated as a surrogate marker representing disease (2–4), since amyloid deposition is related to more rapid progression to dementia (5,6). Amyloid positron emission tomography (PET) and cerebrospinal fluid (CSF) sampling can directly identify the presence of Aβ; however, their limited accessibility and high cost restrict their widespread use.

Amyloid deposition has been implicated in the structural alteration of the brain. Such alterations, manifesting as cerebral atrophy or hippocampal volume loss, can be quantitatively assessed utilizing structural magnetic resonance imaging (MRI) scans (5–8). Furthermore, numerous machine learning-based studies have highlighted the utility of MRI data to predict amyloid status when integrated with clinical and genetic information (9–16). Most of these studies utilized T1w image-based features due to its excellent tissue contrast, resolution, and widespread availability in public datasets, and employed various machine learning methods to predict Aβ positivity. A few of these specifically investigated a deep learning based approach using T1w images, avoiding the use of subjectively determined features (9,15,16). Although T1w images are useful to visualize brain anatomy, other imaging sequences, particularly T2-weighted fluid attenuated inversion recovery (T2-FLAIR), better delineate white matter abnormalities, which may be associated with amyloid deposition or cognitive impairment (17–21). For example, a recent study showed regional associations between white matter hyperintensities (WMH) and amyloid-PET determined amyloid deposition (19). In general, downstream prediction may benefit from multi-contrast MRI inputs, given that different sequences may highlight different aspects of the disease. Finally, none of these previous studies have tested their model's performance in a truly external test set.

In this study, we trained and tested a deep learning algorithm to predict amyloid positivity in a large publicly accessible cohort of patients using T1w MRI and tested it on a separate external test group. We then evaluated whether a multi-contrast approach could enhance model performance, using T1w and T2-FLAIR as inputs to a deep learning-based method.

**Methods**

*Datasets*
In this study, the data were obtained from three largest publicly available datasets and one in-house dataset (hereafter referred to as the Stanford data). The publicly available datasets were Alzheimer's Disease Neuroimaging Initiative (ADNI), Open Access Series of Imaging Studies 3 (OASIS3), and Anti-Amyloid Treatment in Asymptomatic Alzheimer's Disease (A4). Only subjects who underwent T1w, T2-FLAIR, and amyloid PET were included in the study. For the public datasets, MRI studies were paired with corresponding amyloid PET imaging performed within 30 days for each subject. This study included only datasets from the years 2010 to 2023, given changes in MRI and PET technology over the past decade. Regarding the amyloid-PET data, inclusion was limited to datasets acquired using the following US Food and Drug Administration-approved radiotracers: $^{18}$F-florbetapir (FBP) and $^{18}$F-florbetaben (FBB).

The Stanford dataset was evaluated as an external test set. 231 subjects underwent T1w, T2-FLAIR, and amyloid PET scans (all with FBB). The study protocols for the Stanford data were approved by the Stanford University Institutional Review Board, with written informed consent obtained from each study participant or their legally authorized representative. The MRI and PET scans were acquired simultaneously at Stanford University using a PET/MRI scanner (Signa 3T, GE Healthcare). The acquisition parameters of amyloid PET scanning are detailed in a previous study (22). Each participant's cognitive status and etiology were determined by a clinical consensus of a panel of Alzheimer Disease Research Center-affiliated neurologists at Stanford University. Subjects with non-amnestic cognitive impairment or missing cognitive status were excluded from this study. Flow diagrams for both cohorts are shown in Figure 1, while demographic information can be found in Table 1.

(A)

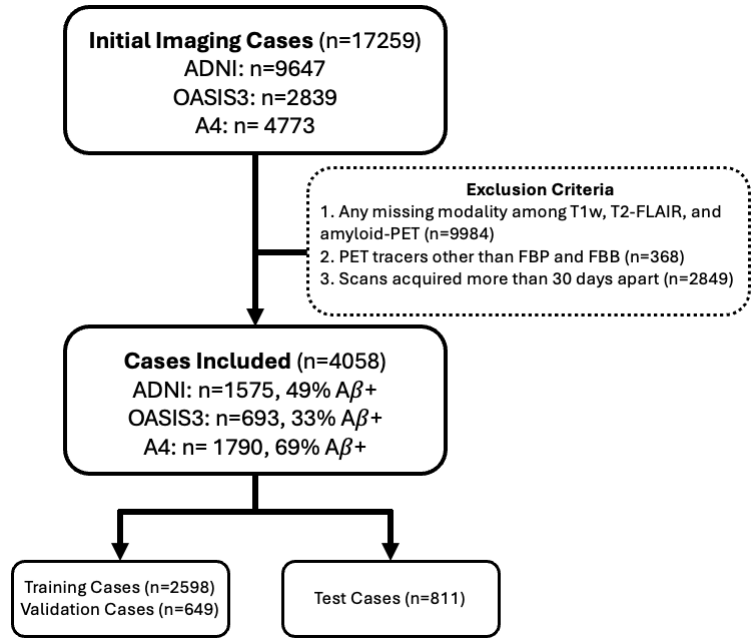

(B)

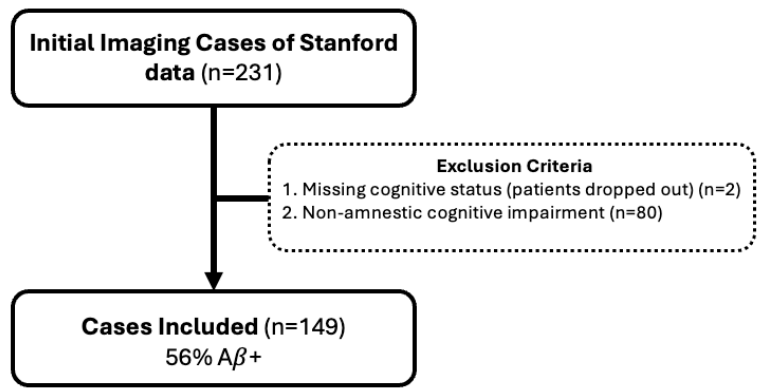

**Figure 1.** Flow diagrams show inclusion of study participants: (A) the public datasets and (B) Stanford dataset.

(A)

| Public | Amyloid Positive (n = 2227) | Amyloid Negative (n = 1831) | P-value |
|---|---|---|---|
| **Age (years)** | 72.6 ± 5.9 | 70.3 ± 6.6 | <0.001 |
| **Sex** | | | 0.025 |
|    Male | 1045 | 794 | |
|    Female | 1182 | 1037 | |
| **Cognitive Status** | | | <0.001 |
|    Normal | 1533 | 1276 | |
|    MCI | 430 | 383 | |
|    Dementia | 153 | 12 | |

(B)

| External | Amyloid Positive (n = 83) | Amyloid Negative (n = 66) | P-value |
|---|---|---|---|
| **Age (years)** | 73.1 ± 9.3 | 70.8 ± 9.9 | 0.153 |
| **Sex** | | | 0.343 |
|    Male | 39 | 25 | |
|    Female | 44 | 41 | |
| **Cognitive Status** | | | <0.001 |
|    Normal | 40 | 55 | |
|    MCI | 18 | 9 | |
|    Dementia | 25 | 2 | |

**Table 1.** Subjects' characteristics at time of amyloid PET imaging: (A) Included cohorts from ADNI, OASIS3, and A4 study; (B) External test set (Stanford data). Values are presented as mean ± s.d. or as the numbers of subjects.

### *MRI Data Acquisition and Post-Processing*
For the Stanford dataset, T1w images were acquired using a sagittal spoiled gradient recalled acquisition with the following parameters: echo time (TE) = 3.06-3.23 ms, repetition time (TR) = 7.65-8.02 ms, inversion time (TI) = 400 ms, and flip angle = 11°. 3D T2-FLAIR images were acquired with the following parameters: TE = 116-165 ms, TR = 4800-6000 ms, TI = 1440-1800 ms, and flip angle = 90°.

T2-FLAIR images underwent co-registration to their corresponding T1w images using rigid registration with Advanced Normalization Tools (ANTS) (23). Brain masks were generated, and images were skull stripped using an artificial neural network-based algorithm (termed HD-BET) on the T1w images (24). All images were oriented in the left-posterior-inferior (LPI) orientation and were resampled using trilinear interpolation to achieve isotropic voxel size of 1 mm. The images were then normalized based on the 5th to 95th intensity percentiles.

### *Centiloid Calculation and Amyloid Status Determination*
Amyloid status was determined based on the centiloid cutoff values recommended for each dataset and their respective radiotracers. For the OASIS3 dataset, this centiloid cutoff value was 20.6, without further details given. The ADNI, A4, and Stanford datasets were processed using an optimized MRI-free Aβ quantification approach (25). A brain atlas and a whole cerebellum mask, obtained from the Global Alzheimer's Association Interactive Network (GAAIN; gaain.org/centiloid-project), were used to calculate standard uptake value ratio (SUVR) values, which were then converted to centiloid values. The ADNI Core developed tracer-specific centiloid cutoff values, which were 12 for FBB and 18 for FBP.

*Deep Learning Model and Experiments*
The public datasets were independently and randomly split into a training set (comprising 64% of the cases), a validation set (comprising 16% of the cases), and a held-out test set (comprising 20% of the cases). The training and validation datasets were used during training, and the held-out test set was used for evaluation. Note that the longitudinal data from the public datasets were used only for training and validation and were excluded from the held-out test set. A 3D EfficientNet-B3 was adopted from the MONAI framework as the deep learning method for this study (26,27). This architecture includes an input layer, multiple convolutional layers, and blocks designed to balance network depth, width, and resolution efficiently. It also features global average pooling to reduce the spatial dimensions before feeding the data into a fully connected layer, which handles the final classification task.

Two separate models were trained for amyloid positivity prediction: one with only T1w images and one with both T1w and T2-FLAIR images as the inputs (Figure 2). All models were trained for 500 epochs with a batch size of 8 with Adam optimizer (28), a cosine annealing learning rate scheduler (29) with a 20-epoch warm up period and a learning rate of 0.0005 after warmup. Binary cross entropy was used as the loss function. The input images were augmented by randomly rotating them along each of the three spatial axes within a range of ±0.2 radians (equivalent to ±11 degrees), with a 30% chance of rotation per axis. 5-fold cross-validation within the training set was performed to enhance the robustness and generalizability of the findings. The output of the model is a number between 0 and 1 representing the likelihood that the scan is amyloid positive. Youden's *J* index (30) was used to define the threshold value. The held-out test set was evaluated by each of these 5 trained model from each fold for the cross-validation.

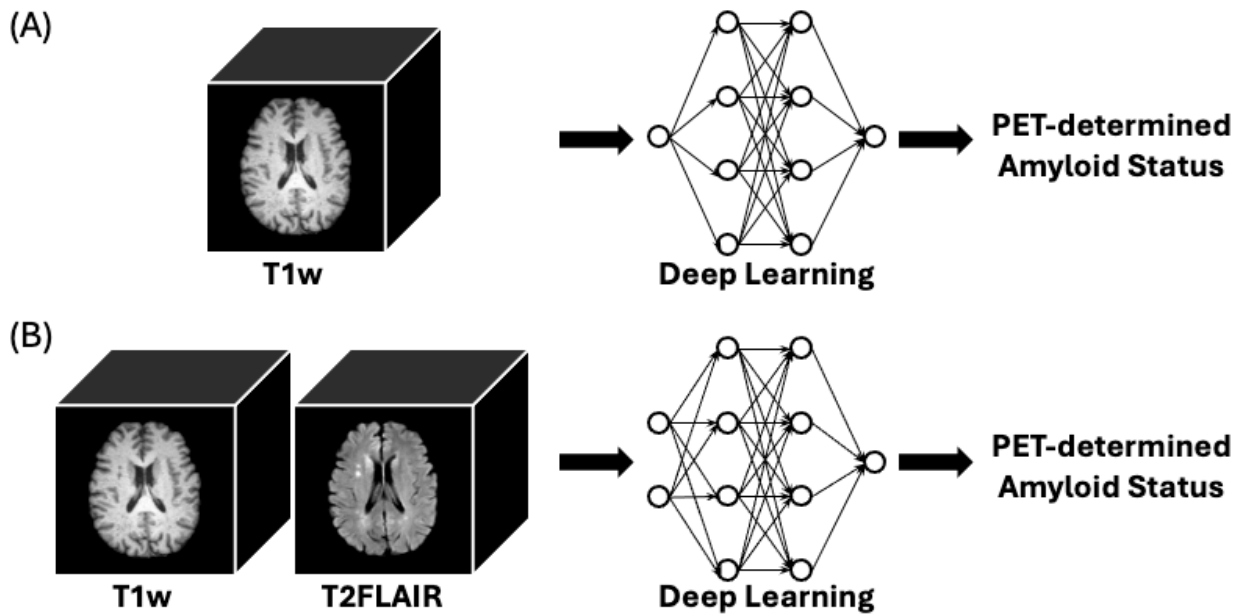

**Figure 2**. Schematic showing the two separate models that were trained for binary amyloid positivity prediction: (A) one with only T1w images, and (B) one with both T1w and T2-FLAIR images as inputs.

Additional analyses were conducted in the held-out internal test set, evaluated both in the entire group as well as in patients with different cognitive statuses, including cognitively normal (CN), mild cognitive impairment (MCI), and dementia, using each of pretrained 5 weights from each model. Note that in these cohorts, the dementia group may include both AD and dementia from other causes (i.e., not all of the dementia cases are amyloid positive).

To further evaluate the generalizability of the models, we tested the trained weights on an external dataset from Stanford University, representing a completely separate imaging protocol and different study population from those used to train the network. Due to the small sample sizes in each cognitive subgroup within this external test set, we did not perform subgroup analyses based on cognitive status.

To identify the key regions influencing classification, we visualized the regional effects of the input images for both the T1w-only and T1w+T2FLAIR models. A sliding 7×7×7 voxel kernel was applied to mask local regions of the input MRI images by setting their voxel values to zero. Each modified image was then processed by the pre-trained model to obtain a prediction score, which was assigned to the central voxel of the masked region, generating an activation map. In this activation map, each voxel's value reflects the model's prediction score when that region was masked, indicating its impact on the prediction. The activation map was normalized and adjusted to represent to absolute change, so that lower scores indicate that the masked region had a greater impact on the model's prediction.

### *Comparison with a Prior Model*
A previous study developed a deep learning model that predicts amyloid status using T1w images alone (9). This model had significantly fewer free parameters compared to the model in the current study, with only 5 convolutional layers, compared to 131 convolutional layers in the current model. For comparative analysis, we utilized their pretrained weights to test the model's performance on our held-out test set within the ADNI cohort, following their preprocessing steps, as their model was originally developed using ADNI data. Additionally, we re-trained their model using our own training and validation datasets, and then evaluated its performance on our held-out test set as well as the external test set. Given that this model was designed to work solely with T1w images, we could not develop a T1w+T2-FLAIR version based on their model. We assessed the model's performance by calculating AUC, accuracy, sensitivity, and specificity across the validation set, held-out test set, and external test set.

### *Statistical evaluation and Performance Evaluation*
Study subjects' demographics were stratified by amyloid-beta status into Aβ+ and Aβ- groups. Predictions from the five-fold cross-validation were aggregated for performance evaluation and statistical analysis. The demographic characteristics were compared using chi-squared tests and t-tests for categorical variables and quantitative variables, respectively. The primary performance metric was the area under the receiver operating characteristics curve (AUC). To compare the performance of T1w-only and T1w+T2-FLAIR models, DeLong's test and McNemar's test were used to compare the AUCs and accuracies, respectively. Binary predictions were assessed using accuracy, sensitivity, and specificity using Youden's *J* index.

**Results**

### *Study Sample Characteristics*
Figure 1 presents flowcharts illustrating the selection process of study subjects, starting from the initial number of imaging participants, followed by the application of specific exclusion criteria, and resulting in the final cohort used for training or validation. For network development (Figure 1-(A)), this study included a total of 4,058 subjects with T1w and T2-FLAIR MRI scans from the public datasets, along with their amyloid PET-determined statuses, of which 55% were Aβ+. The characteristics of these subjects included are summarized in Table 1-(A). There were significant differences between the Aβ+ and Aβ- groups in terms of age, sex distribution, and cognitive status, with the amyloid positive subjects being more likely to be older, male, and to have MCI or dementia. A total of 149 subjects were included in the external test set (Figure 1-(B)), and their characteristics are summarized in Table 1-(B). In the external test set, 56% of the subjects were Aβ+, and the Aβ+ subjects were more likely to have MCI or dementia. Figure 3 presents the distributions of centiloid values for selected samples from public and Stanford datasets.

(A)

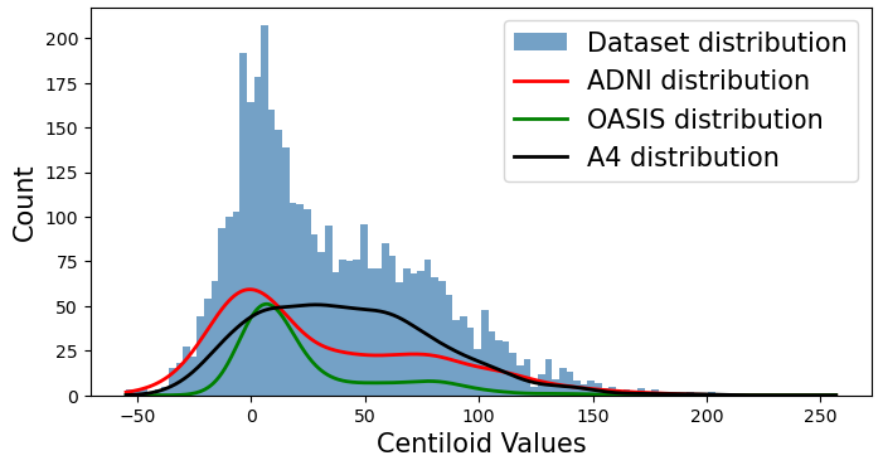

(B)

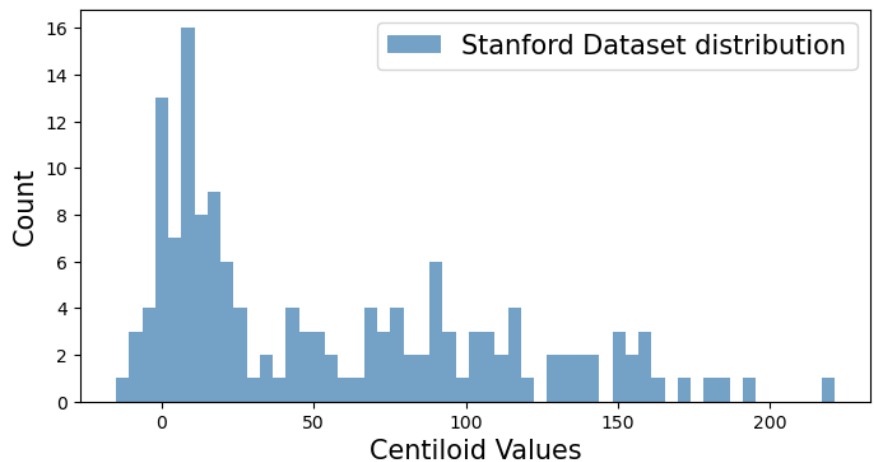

**Figure 3.** Centiloid distribution of the included study cohort: (A) the public datasets and (B) the Stanford dataset.

*Prediction Performance in the Validation and Internal Test Sets*
In both the validation and held-out internal test sets, models incorporating T2-FLAIR in addition to T1w images provided superior performance metrics (Table 2-(A)). Specifically, in the held-out internal test set, the prediction of amyloid positivity was as follows: AUCs were 0.62 (95% CI: 0.60 and 0.64) and 0.67 (95% CI: 0.64 and 0.70), accuracies were 61% (95% CI: 60% and 63%) and 64% (95% CI: 62% and 66%), sensitivities were 0.88 (95% CI: 0.87 and 0.89) and 0.71 (95% CI:

0.70 and 0.73), and specificities were 0.23 (95% CI: 0.22 and 0.24) and 0.53 (95% CI: 0.52 and 0.55) for the T1w-only and T1w+T2-FLAIR models, respectively. These differences were statistically significant, as indicated by DeLong's test (p=0.006) and McNemar's test (p=0.008). The receiver operating characteristic (ROC) curves from the validation and held-out test sets are shown in Figure 4. Youden's *J* index values were found to be 0.11 and 0.25 from the validation data for T1w-only model and the T1w+T2-FLAIR model, respectively.

(A)

|  | Validation Set | | Internal Test Set | |
|---|---|---|---|---|
|  | T1w only | T1w & T2FLAIR | T1w only | T1w & T2FLAIR |
| **All** | | | | |
| AUC | 0.63 (0.60, 0.65) | 0.70 (0.67, 0.73) | 0.62 (0.60, 0.64) | 0.67 (0.64, 0.70) |
| DeLong's test | P<0.001 | | P=0.006 | |
| Accuracy | 0.62 (0.60, 0.63) | 0.66 (0.64, 0.68) | 0.61 (0.60, 0.63) | 0.64 (0.62, 0.66) |
| McNemar's test | P<0.001 | | P=0.008 | |
| Sensitivity | 0.89 (0.87, 0.90) | 0.73 (0.72, 0.75) | 0.88 (0.87, 0.89) | 0.71 (0.70, 0.73) |
| Specificity | 0.23 (0.22, 0.25) | 0.56 (0.54, 0.58) | 0.23 (0.22, 0.24) | 0.53 (0.52, 0.55) |

(B)

|  | External Test Set | |
|---|---|---|
|  | T1w only | T1w & T2FLAIR |
| **All** | | |
| AUC | 0.55 (0.50, 0.60) | 0.65 (0.59, 0.70) |
| DeLong's test | P=0.014 | |
| Accuracy | 0.54 (0.50, 0.57) | 0.60 (0.57, 0.64) |
| McNemar's test | P=0.004 | |
| Sensitivity | 0.74 (0.71, 0.77) | 0.78 (0.75, 0.81) |
| Specificity | 0.31 (0.27, 0.34) | 0.39 (0.36, 0.43) |

**Table 2.** Performance metrics for amyloid positivity prediction from (A) internal validation and held-out test set, and (B) the external test set. Values in parentheses represent 95% confidence intervals. Accuracy, sensitivity, and specificity metrics are presented for a threshold representing the Youden's *J* index operating points.

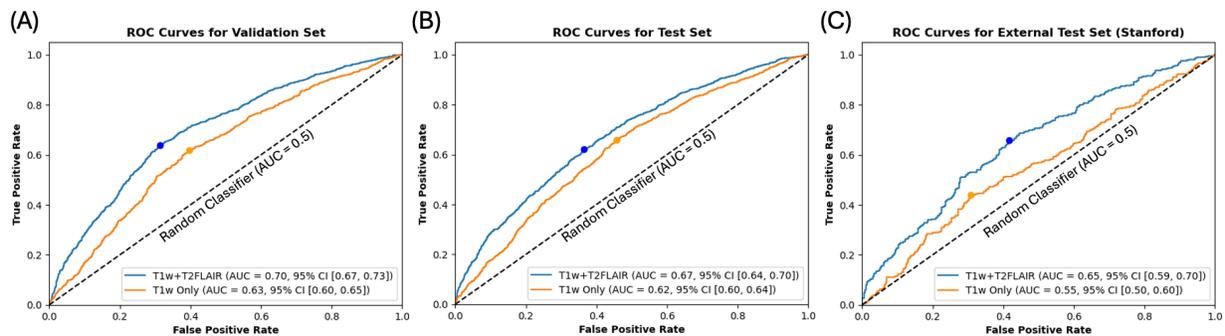

**Figure 4**. ROC curves of the T1w-only and T1w+T2-FLAIR models in the (A) validation set, (B) internal held-out test set, and (C) external test set. In all settings, adding T2-FLAIR imaging improves performance. The blue and yellow scatter points represent the optimal cut-off points for the T1w+T2-FLAIR and T1w-only models, respectively.

*Prediction by Different Cognitive Status*
In the analysis across different cognitive statuses, the T1w-only model provided moderate AUC values in the CN and MCI groups, which were 0.61 (95% CI: 0.58 to 0.64) and 0.68 (95% CI: 0.62 to 0.73), respectively. However, the dementia group showed an AUC of 0.47 (95% CI: 0.33 to 0.62). The addition of T2-FLAIR images resulted in higher AUC and accuracy across all cognitive groups, as shown in Table 3. Notably, the highest AUC value was observed in the MCI group for the T1w+T2-FLAIR model, with an AUC of 0.71 (95% CI: 0.64 to 0.79).

| | Internal Test Set | |
|---|---|---|
| | T1w only | T1w & T2FLAIR |
| **CN** (n=502, 59% A$\beta$+) | | |
| AUC | 0.61 (0.58, 0.64) | 0.66 (0.63, 0.70) |
| DeLong's test | P=0.018 | |
| Accuracy | 0.61 (0.59, 0.63) | 0.64 (0.62, 0.65) |
| McNemar's test | P=0.025 | |
| Sensitivity | 0.88 (0.86, 0.89) | 0.72 (0.70, 0.74) |
| Specificity | 0.23 (0.22, 0.25) | 0.52 (0.50, 0.53) |
| **MCI** (n=97, 55% A$\beta$+) | | |
| AUC | 0.68 (0.62. 0.73) | 0.71 (0.64. 0.79) |
| DeLong's test | P=0.435 | |
| Accuracy | 0.60 (0.55, 0.64) | 0.64 (0.60, 0.69) |
| McNemar's test | P=0.092 | |
| Sensitivity | 0.91 (0.88, 0.93) | 0.73 (0.69, 0.77) |
| Specificity | 0.22 (0.19, 0.26) | 0.54 (0.49, 0.58) |
| **Dementia** (n=31, 94% Aβ+) | | |
| AUC | 0.47 (0.33, 0.62) | 0.61 (0.35, 0.88) |
| DeLong's test | P=0.366 | |
| Accuracy | 0.86 (0.81, 0.92) | 0.69 (0.62, 0.76) |
| McNemar's test | P<0.001 | |
| Sensitivity | 0.92 (0.88, 0.97) | 0.72 (0.65, 0.79) |
| Specificity | 0 (0, 0) | 0.30 (0.23, 0.37) |

**Table 3.** Performance metrics for amyloid positivity prediction by cognitive status in the internal held-out test sets. Values in parentheses represent 95% confidence intervals. Accuracy, sensitivity, and specificity metrics are presented for a threshold representing the Youden's *J* index operating points.

### *Prediction on An External Test Set*

Evaluation of the trained weights on the external test set demonstrated prediction performance comparable to that of the held-out test set used in this study, as shown in Table 2. Consistent with our findings in the internal held-out test set, the T1w+T2-FLAIR model outperformed the T1w-only model in the external test set, showing superior performance in terms of AUC, accuracy, sensitivity, and specificity (Table 2-(B) and Figure 4-(C)). Moreover, the differences in AUC and accuracy were statistically significant, as confirmed by DeLong's test (p = 0.014) and McNemar's test (p = 0.004) (Table 2-(B)). These results suggest that our models, trained on publicly available datasets, can be effectively applied to data from a different center and population.

### *Visualization of Regional Effect*

Figure 5 illustrates activation maps showing the regional effects of the input images for randomly selected Aβ+ and Aβ- subjects. Both T1w-only and T1w+T2FLAIR models exhibited different activation patterns in the gray and white matter for Aβ+ and Aβ- cases. Notably, in the T1w+T2FLAIR model, regions surrounding the brain's ventricles displayed stronger activations compared to the T1w-only model. This observation aligns with the fact that periventricular WMH are commonly detected in these regions. Additionally, Figure 6 presents the T1w, T2-FLAIR, and amyloid PET images corresponding to the maximum and minimum predicted scores of the T1w+T2FLAIR model from the held-out test set. The case with the maximum predicted score showed posterior cortical atrophy and severe white matter hyperintensities, in contrast to the case with the minimum predicted score.

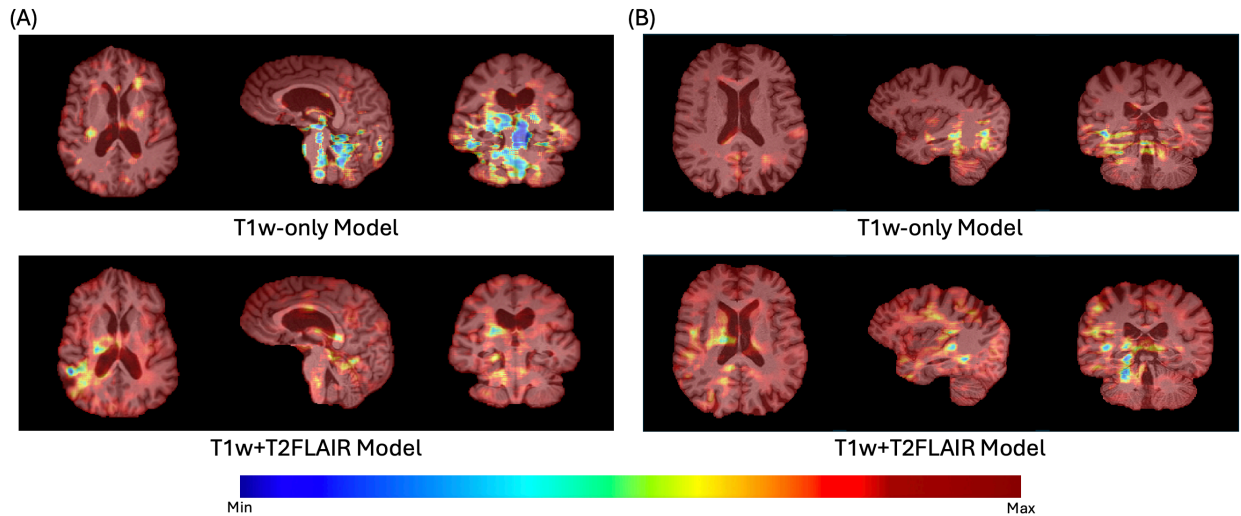

**Figure 5**. Activation maps showing the regional effects of the input images for randomly selected (A) Aβ+ and (B) Aβ- subjects. The upper and lower images show the maps from T1w-only and T1w+T2FLAIR models, respectively.

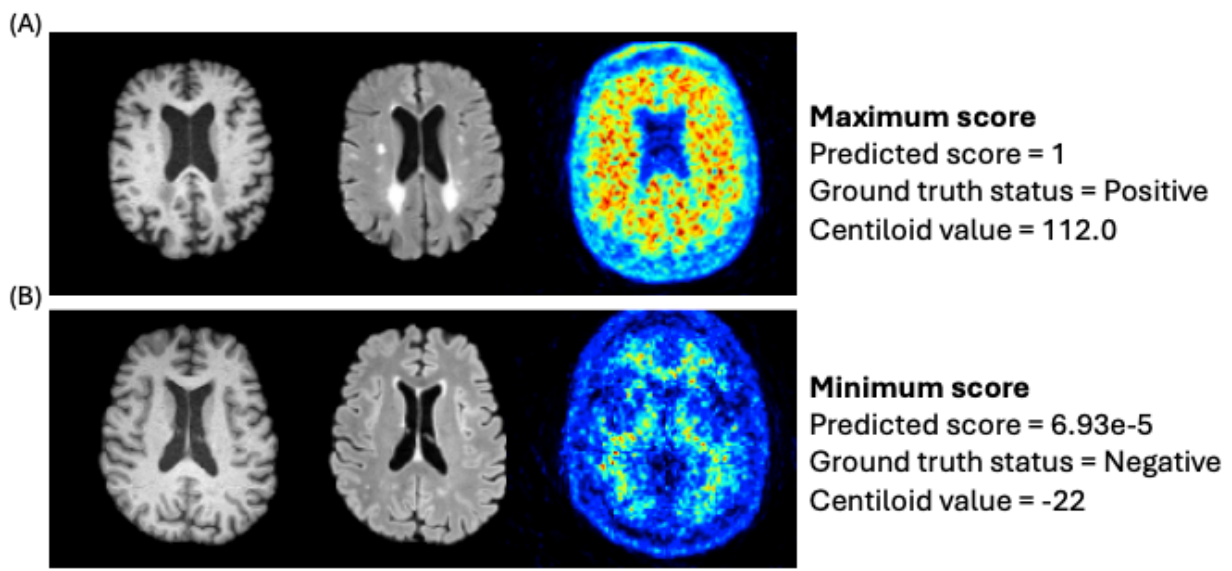

**Figure 6.** The T1w, T2-FLAIR, and amyloid PET images of (A) the maximum and (B) the minimum predicted scores from T1w+T2-FLAIR model.

*Comparison with A Prior Model*

The pretrained weights from the previous study (9) were not able to successfully predict the amyloid status of our ADNI dataset, demonstrating an AUC of 0.57, accuracy of 48%, sensitivity of 0.95, and specificity of 0.07. Therefore, their model was re-trained using our own training and validation datasets. The prediction results for the retrained model for our validation, internal held-out test set, and external test set are presented in Table 4 with Youden's *J* index value of 0.5. The model provided similar prediction performance of our T1w-only model in AUC and accuracy. However, the prior model was not able to predict the external test set. Additionally, Figure 7 shows the histograms of the predicted scores of our models and the prior model in the held-out test set.

While the current model demonstrated predictions across the entire likelihood range (0-1), the prior model provided a score near 0.5 for most cases (Figure 7-(B)).

|  | Validation Set (n = 649) | Internal Test Set (n = 811) | External Test Set (n = 149) |
|---|---|---|---|
| AUC | 0.61 (0.61, 0.62) | 0.61 (0.61, 0.62) | 0.45 (0.44, 0.46) |
| Accuracy | 0.57 (0.55, 0.59) | 0.60 (0.58, 0.62) | 0.46 (0.42, 0.49) |
| Sensitivity | 0.52 (0.50, 0.54) | 0.67 (0.65, 0.68) | 0.41 (0.37, 0.44) |
| Specificity | 0.65 (0.63, 0.66) | 0.50 (0.49, 0.52) | 0.52 (0.49, 0.56) |

**Table 4.** Performance metrics for amyloid positivity prediction using the prior model on the internal validation, internal held-out test, and external test sets are provided. The results are shown for the T1w-only model, as the prior model can only use T1w images as input.

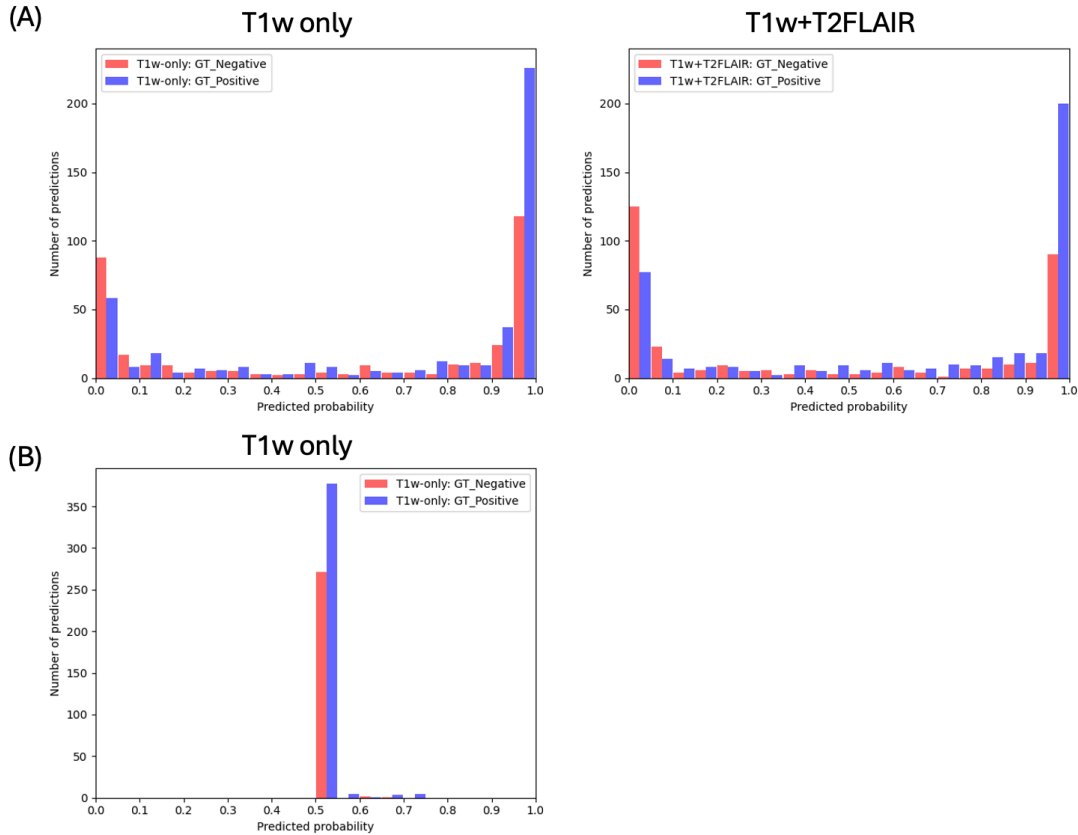

**Figure 7.** Histograms of the predicted scores from the current model and a previously published model (9). For the current model: (A) the T1w-only model (left) and the T1w+T2FLAIR model (right). (B) The previous model. Red and blue bars represent the ground truth statuses (GT: Negative and GT: Positive, respectively). Each pair of adjacent red and blue bars corresponds to the same bin on the x-axis.

## Discussion

Amyloid PET is currently considered one of the gold standards for *in vivo* imaging to detect amyloid protein in the brain. Despite its precision, the technique has drawbacks such as limited accessibility, high costs, and the need for invasive procedures. In our study, we sought to estimate amyloid status, as determined by PET, using readily accessible MRI images alone, without requiring hand-crafted features such as cortical segmentations. We specifically utilized a deep learning model that incorporated T1w and T2-FLAIR images. This model achieved an AUC of 0.67, accuracy of 64%, sensitivity of 0.71, and specificity of 0.53. We observed notable performance improvement with the addition of T2-FLAIR volumes in the validation, held-out internal test, and external test sets. These findings suggest that adding multi-modal MR information in the form of T2-FLAIR images significantly enhances the ability of the model to distinguish amyloid status.

Overall, we observed prediction at better than chance levels, but still with AUC's in the 0.6-0.7 range, which reflects only moderate performance. There are several probable reasons for this, the most important of which is that amyloid changes are known to precede major structural changes in the brain by decades (31). The ability to predict at better than chance across the different cognitive status classes suggests that there are informative features. However, very early cases of amyloid positivity may not show sufficient MRI changes for prediction, for example. Another possibility is that the size of the dataset used, though quite large by neuroimaging study standards, is still too small to extract the full information from the scans. Nevertheless, given the non-invasiveness and availability of MRI, such performance could be useful for clinical trial screening or to identify potential at-risk individuals using a high specificity threshold using opportunistic screening.

T2-FLAIR images are obtained in imaging protocols for most diseases, including studies acquired for memory concerns. Unlike T1w images, they highlight regions of abnormality with positive contrast, making them sensitive to detect subtle pathology, including WMH that are thought to represent small vessel ischemic disease in elderly subjects. The prediction results across different cognitive statuses demonstrated a clear advantage when T2-FLAIR images were added to T1w images, regardless of cognitive status. The reason for the improvement could relate to the better delineation of WMH, as a previous study showed regional associations between WMH and Aβ accumulation (19). However, simply having additional imaging data, resulting in better visualization of brain structure regardless of the different contrast, could also be a reason for the generally improved performance. Additionally, the activation maps revealed that the T1w+T2FLAIR model exhibited stronger activations in regions surrounding the brain's ventricles, areas particularly vulnerable to periventricular WMH (Figure 5).

Previous studies have demonstrated the potential of using MRI data to predict amyloid status as determined by amyloid PET scans through machine learning methods (9–16). The majority of these studies primarily utilized T1w images. Most of the previous works used hand-crafted features such as segmented cortical volumes as inputs to the machine learning model, a step that can be prone to error and require manual correction. Additionally, many of these previous studies also used demographic, genetic, and clinical information, which is not always available, and which may be subjective, such as distinguishing between MCI and dementia.

In our study, we excluded any demographic, genetic, or clinical information, so as to focus only on the information in the image and the impact of using multiple contrasts. Our model using only T1w images yielded results comparable to those in a previous whole-image deep learning study

(9). Interestingly, in this previous work, the performance in the validation set (AUC of 0.63) was significantly lower than in the test set (AUC of 0.73), raising questions about the composition of the test set given that it is unusual for performance in the test set to exceed that in the validation set for deep learning studies. Furthermore, we could not reproduce similar results using their pretrained weights, despite the fact that our held-out internal test set may have included examinations from their training set (which should have inflated performance). Additionally, when we re-trained the previous model on our training dataset, the newly trained weights were not able to predict our external test set effectively, achieving an AUC of only 0.45 ([Table 4](#)). Other previous studies that utilized MRI volumes reported an AUC of 0.74 (15) and balanced accuracy of 0.76 (16), with limited details provided. However, all of these previous studies (9,15,16) used only T1w images from the ADNI cohort. Therefore, the use of T2-FLAIR images remains relatively unexplored. Although some studies have incorporated hand-crafted imaging features derived from T2-FLAIR (10,32), such as the volume of white matter hyperintensities, none have directly employed T2-FLAIR volumes themselves. Also, this study is the first to report performance in an external test set, highlighting the model's robustness and generalizability across different populations and imaging protocols. Moreover, our study showed that adding T2-FLAIR images improved performance in both the internal and external test sets.

There are several limitations to this study. First, we defined amyloid positivity using tracer-specific centiloid threshold values. The centiloid is a standardized scale designed to measure amyloid deposition across various tracers from amyloid PET images. For clinical trials purposes, often the amyloid status is determined human visual reads. Second, this study does not specifically evaluate how T2-FLAIR images improve the MRI-based prediction of PET-determined amyloid status. Although it is hypothesized that information about white matter abnormalities from T2-FLAIR contributes to more accurate amyloid status estimations, the precise mechanisms are not fully explained. Based on the comparison between the maximum and minimum predicted score cases ([Figure 6](#)), the network seems capable of capturing brain structural abnormalities. We have not employed saliency maps, as used in prior works (9), given the growing understanding of their limitations with respect to variability and explainability (33,34). However, we provided activation maps that illustrate the regions of interest that most significantly contribute to the model's decision-making process. These maps were generated by masking local regions of the input images and analyzing the changes in the model's output, allowing for a more intuitive understanding of which areas in the brain images are considered important for classification. Also, we have not added demographic and clinical variables into the model, suggesting that the performance presented represents a lower bound. Multiple prior studies have shown that adding this information can improve predictions (9–12,14,15). However, we chose to examine the relatively simpler question of whether the addition of another imaging sequence with different contrast improved performance; there are many methods to incorporate clinical with imaging data, and it is possible that the addition of optimally chosen and integrated clinical and demographic data might reduce the relative value of the additional T2-FLAIR images.

In conclusion, we found that deep learning models with volumetric imaging data only as inputs can be used to predict amyloid status with about 60-70% accuracy. Incorporating T2-FLAIR images with T1w significantly improved the prediction of PET-determined amyloid status, and this generalized to an external data set. Future work will explore whether additional MRI sequences may further enhance prediction by providing additional information relevant to amyloid deposition as well as whether this additional imaging continues to provide benefit when clinical and demographic features are added to the model.

**Authors' Contributions**
Donghoon Kim: methodology, investigation, writing

Jon André Ottesen: methodology, investigation
Ashwin Kumar: investigation
Brandon Ho: support
Elsa Bismuth: support
Christina B. Young: resources
Elizabeth Mormino: resources
Greg Zaharchuk: design, methodology, supervision, funding acquisition

**Acknowledgements**
ADNI: Data collection and sharing for this project was funded by the Alzheimer's Disease Neuroimaging Initiative (ADNI) (National Institutes of Health Grant U01 AG024904) and DOD ADNI (Department of Defense award number W81XWH-12-2-0012). ADNI is funded by the National Institute on Aging, the National Institute of Biomedical Imaging and Bioengineering, and through generous contributions from the following: AbbVie, Alzheimer's Association; Alzheimer's Drug Discovery Foundation; Araclon Biotech; BioClinica, Inc.; Biogen; Bristol-Myers Squibb Company; CereSpir, Inc.; Cogstate; Eisai Inc.; Elan Pharmaceuticals, Inc.; Eli Lilly and Company; EuroImmun; F. Hoffmann-La Roche Ltd and its affiliated company Genentech, Inc.; Fujirebio; GE Healthcare; IXICO Ltd.;Janssen Alzheimer Immunotherapy Research & Development, LLC.; Johnson & Johnson Pharmaceutical Research & Development LLC.; Lumosity; Lundbeck; Merck & Co., Inc.;Meso Scale Diagnostics, LLC.; NeuroRx Research; Neurotrack Technologies; Novartis Pharmaceuticals Corporation; Pfizer Inc.; Piramal Imaging; Servier; Takeda Pharmaceutical Company; and Transition Therapeutics. The Canadian Institutes of Health Research is providing funds to support ADNI clinical sites in Canada. Private sector contributions are facilitated by the Foundation for the National Institutes of Health (www.fnih.org). The grantee organization is the Northern California Institute for Research and Education, and the study is coordinated by the Alzheimer's Therapeutic Research Institute at the University of Southern California. ADNI data are disseminated by the Laboratory for Neuro Imaging at the University of Southern California.

OASIS-3: Data were provided by OASIS (OASIS-3, Longitudinal Multimodal Neuroimaging: Principal Investigators: T. Benzinger, D. Marcus, J. Morris, were supported by NIH Grants: NIH P30 AG066444, P50 AG00561, P30 NS09857781, P01 AG026276, P01 AG003991, R01 AG043434, UL1 TR000448, and R01 EB009352).

A4: The A4 Study is a secondary prevention trial in preclinical Alzheimer's disease, aiming to slow cognitive decline associated with brain amyloid accumulation in clinically normal older individuals. The A4 Study is funded by a public-private-philanthropic partnership, including funding from the National Institutes of Health-National Institute on Aging, Eli Lilly and Company, Alzheimer's Association, Accelerating Medicines Partnership, GHR Foundation, an anonymous foundation and additional private donors, with in-kind support from Avid and Cogstate. The companion observational Longitudinal Evaluation of Amyloid Risk and Neurodegeneration (LEARN) Study is funded by the Alzheimer's Association and GHR Foundation. The A4 and LEARN Studies are led by Dr. Reisa Sperling at Brigham and Women's Hospital, Harvard Medical School and Dr. Paul Aisen at the Alzheimer's Therapeutic Research Institute (ATRI), University of Southern California. The A4 and LEARN Studies are coordinated by ATRI at the University of Southern California, and the data are made available through the Laboratory for Neuro Imaging at the University of Southern California. The participants screening for the A4 Study provided permission to share their de-identified data in order to advance the quest to find a successful treatment for Alzheimer's disease. We would like to acknowledge the dedication of all the participants, the site personnel, and all of the partnership team members who continue to make the A4 and LEARN Studies possible. The complete A4 Study Team list is available on: www.actcinfo.org/a4-study-team-lists.